\documentclass[conference]{IEEEtran}
\IEEEoverridecommandlockouts
\usepackage{cite}
\usepackage{amsmath,amssymb,amsfonts}
\usepackage{graphicx}
\usepackage{textcomp}
\usepackage{xcolor}
\usepackage{algorithm}
\usepackage{algpseudocode}
\usepackage{booktabs}
\usepackage{multirow}
\usepackage{balance}
\usepackage{url}

\def\BibTeX{{\rm B\kern-.05em{\sc i\kern-.025em b}\kern-.08em
    T\kern-.1667em\lower.7ex\hbox{E}\kern-.125emX}}
\begin{document}

\title{Lightweight Diffusion Models for Resource-Constrained Semantic Communication
\thanks{This work was supported by the European Union under the Italian National Recovery and Resilience Plan (PNRR) of NextGenerationEU, partnership on
``Telecommunications of the Future” (PE00000001 - program RESTART).}
}

\author{\IEEEauthorblockN{Giovanni Pignata, Eleonora Grassucci, Giordano Cicchetti, Danilo Comminiello}
\IEEEauthorblockA{\textit{Department of Information Engineering, Electronics, and Telecommunication, Sapienza University of Rome, Italy} \\
eleonora.grassucci@uniroma1.it}
}

\maketitle

\begin{abstract}
Recently, generative semantic communication models have proliferated as they are revolutionizing semantic communication frameworks, improving their performance, and opening the way to novel applications. Despite their impressive ability to regenerate content from the compressed semantic information received, generative models pose crucial challenges for communication systems in terms of high memory footprints and heavy computational load. In this paper, we present a novel Quantized GEnerative Semantic COmmunication framework, Q-GESCO. The core method of Q-GESCO is a quantized semantic diffusion model capable of regenerating transmitted images from the received semantic maps while simultaneously reducing computational load and memory footprint thanks to the proposed post-training quantization technique. Q-GESCO is robust to different channel noises and obtains comparable performance to the full precision counterpart in different scenarios saving up to 75\% memory and 79\% floating point operations. This allows resource-constrained devices to exploit the generative capabilities of Q-GESCO, widening the range of applications and systems for generative semantic communication frameworks.
The code is available at \url{https://github.com/ispamm/Q-GESCO}.
\end{abstract}

\begin{IEEEkeywords}
Generative Semantic Communication, Quantization, Diffusion Models, Wireless Communication
\end{IEEEkeywords}

\section{Introduction}


In the last year, generative models have profoundly revolutionized semantic communication systems. Relying on the transmission and regeneration of the semantic information between sender and receiver \cite{Qin2024Proceeidngsieee}, such semantic frameworks have been the fertile terrain for generative models tailored for semantic communications to grow \cite{Grassucci2023EnhancingSC, grassucci2024generative, Barbarossa2023COMMAG}.
Indeed, the quality of their generated multimedia content is exceptionally good and realistic. Moreover, generative models excel in regenerating content under the guidance of extremely compressed semantic information, which may be various, ranging from latent vectors to textual captions, audio information, or semantic maps. Such flexibility brought to the definition of a variety of generative semantic communication (GSC) frameworks for image \cite{Grassucci2023GenerativeSC, Nam2024ICASSP, Cicchetti2024LanguageOrientedSL, Guo2024DiffusionDrivenSC}, audio \cite{Grassucci2024GSC, Tian2024ICC}, or scene transmission \cite{Yang2024ICASSP, Jiang2024ICASSP} that outperform classical communication methods, opening the path to more bandwidth-efficient and expressive communication frameworks \cite{grassucci2024rethinking, grassucci2024generative}.

Despite this revolution, the computational demands of generative models pose significant challenges for the effective deployment of GSC frameworks, particularly in resource-constrained environments such as mobile devices and edge computing platforms \cite{Lai2024TWC, Wang2024TMC}. Diffusion models, which are the current state-of-the-art methods for generative modelling \cite{Dhariwal2021DiffusionMB}, transform real data gradually into Gaussian noise, and then reverse the process to generate real data from Gaussian noise. Therefore, the process is iterative and may require numerous forward passes through the model to generate a single output. This not only increases the inference time but also demands substantial memory resources, which can be prohibitive for deployment in real-time or low-power applications. Moreover, the large size of diffusion models, often comprising millions of parameters, exacerbates these challenges, making them impractical for scenarios where computational efficiency and latency are critical, as in communication \cite{xu2024semanticaware, Qiao2024LatencyAwareGS}.

A widely adopted technique to reduce the computational burden of deep learning models is quantization, which has gained attention as a promising solution to address the inefficiencies associated with diffusion models \cite{Nagel2020UpOD, Li2021BRECQPT}. By reducing the precision of model parameters, quantization significantly lowers both the memory footprint and the computational load, enabling faster inference and reduced power consumption \cite{li2023qdiffusionquantizingdiffusionmodels, shang2023posttrainingquantizationdiffusionmodels}.

\begin{figure*}
    \centering
    \includegraphics[width=0.99\textwidth]{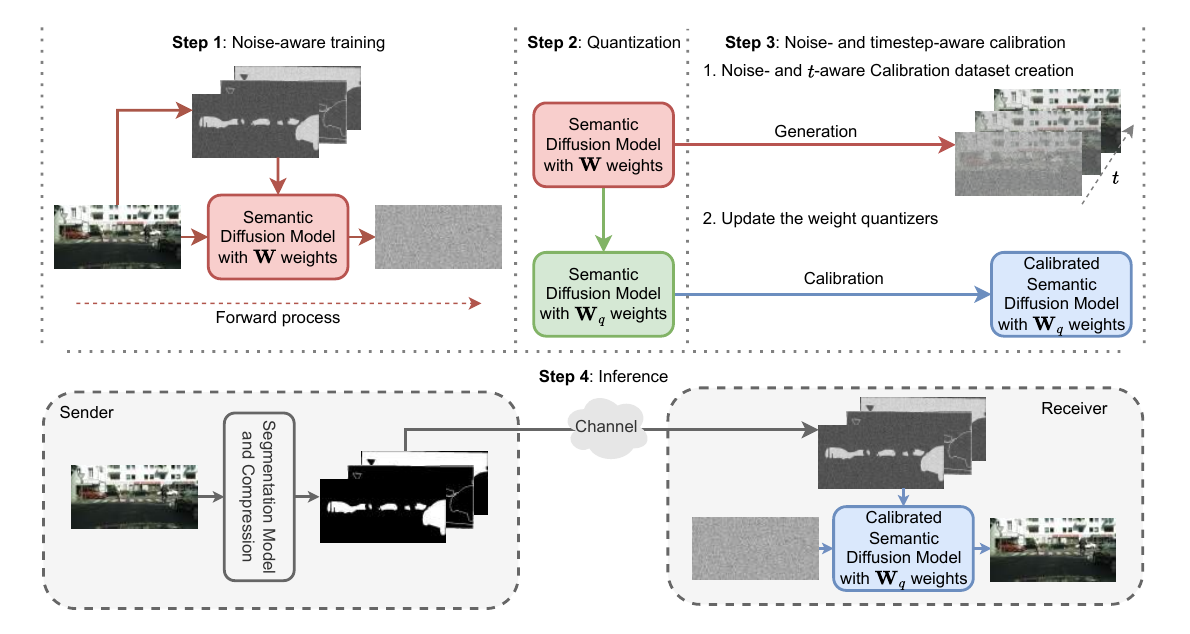}
    \caption{Q-GESCO pipeline.}
    \label{fig:method}
\end{figure*}

Encouraged by the promising results of diffusion models quantization and generative semantic communication frameworks, in this paper, we introduce Q-GESCO: Quantized-GEnerative Semantic COmmunication framework, an effective yet lighter diffusion model for semantic communication in resource-constrained scenarios. The core of Q-GESCO is at the receiver, consisting of a semantic diffusion model that guides the regeneration process by means of the semantic maps transmitted from the sender over the communication channel. The post-training quantization (PTQ) of Q-GESCO allows significant compression of the denoising neural network inside the diffusion model in a completely data-free way, while preserving the generative performance of the method, comparable to the full precision counterpart. In this way, Q-GESCO is capable of: i) regenerating high-quality and semantic-rich image content which is barely affected by the channel noise, ii) reducing the bandwidth required for the transmission by sending highly-compressed semantic maps, and iii) lowering the computational load and the memory footprint of the network in the semantic diffusion model, enabling the implementation of Q-GESCO in resource-constrained environments.

The rest of the paper is organized as follows: Section~\ref{sec:method} introduces the core methods of Q-GESCO, experiments are conducted in Section~\ref{sec:exp} and conclusions are drawn in Section~\ref{sec:con}.

\section{Quantizing Diffusion Models for Semantic Communication} 
\label{sec:method}

In this work, we propose the Quantized GEnerative Semantic COmmunication (Q-GESCO) framework that relies on post-training quantization (PTQ) techniques for the semantic diffusion model. Specifically, we focus on GESCO \cite{Grassucci2023GenerativeSC}, the state-of-the-art framework developed for semantic communication based on diffusion models. While GESCO has demonstrated strong performance in generating semantically accurate image content, it demands high computational resources to be practically run in real-world scenarios. However, communication systems usually work with resource-constrained devices, and the inference of these huge and demanding models may be impractical in such devices. To this end, we propose Q-GESCO, which crucially reduces the computational load and the memory footprint of GESCO through post-training quantization and ad-hoc calibration techniques while inheriting its generative performance.

The Q-GESCO pipeline can be visualized in Fig.~\ref{fig:method} and formalized as follows:

\noindent \textbf{Step 1: Noise-aware training.} Full precision training (or getting pretrained) noise-aware GESCO on the desired dataset, set different noise values to apply to the semantic maps so as to emulate the channel noise.

\noindent \textbf{Step 2: Quantization.} Quantize the trained GESCO with post-training quantization to obtain the quantized model Q-GESCO.

\noindent \textbf{Step 3: Noise- and timestep-aware calibration.} Create the calibration dataset (noise- and timestep-aware), then calibrate the quantized weights of Q-GESCO.

\noindent \textbf{Step 4: Inference.} Run Q-GESCO in semantic communication scenarios by transmitting compressed semantic maps with low bandwidth and regenerating high-quality images in resource-constrained devices.

As emerges from the pipeline, Q-GESCO has several sub-modules that we present in the next subsections.

\subsection{Semantic Diffusion Model}

The core structure of Q-GESCO is a semantic diffusion model (SDM) designed to incorporate semantic conditioning into the diffusion process to guide the generation process preserving the semantics of the transmitted content.
Given an input sample $\mathbf{x}_0$, the forward process gradually adds Gaussian noise to the data over $T$ timesteps, yielding a sequence of noisy content $\mathbf{x}_1, \mathbf{x}_2, \dots, \mathbf{x}_T$. This process is governed by the following transition probability:

\begin{equation}
q(\mathbf{x}_t | \mathbf{x}_{t-1}) = \mathcal{N}(\mathbf{x}_t; \sqrt{\alpha_t} \mathbf{x}_{t-1}, (1 - \alpha_t)\mathbf{I}),
\end{equation}

\noindent where $\alpha_t$ are noise scaling coefficients.

In the reverse process, the model progressively denoises the noise sample $\mathbf{x}_T$, generating a sequence of samples up to the data content $\mathbf{x}_0$. The reverse process is parameterized by a denoising network $\epsilon_\theta(\mathbf{x}_t, t, \mathbf{y})$, conditioned on the semantic maps $\mathbf{y}$, and is defined as:

\begin{equation}
p_\theta(\mathbf{x}_{t-1} | \mathbf{x}_t, \mathbf{y}) = \mathcal{N}(\mathbf{x}_{t-1}; \mu_\theta(\mathbf{x}_t, t, \mathbf{y}), \sigma^2_{\theta,t} \mathbf{I}),
\end{equation}

\noindent where $\mu_\theta(\mathbf{x}_t, t, \mathbf{y})$ is the predicted mean and $\sigma_{\theta,t}^2$ is the variance learned by the model.





The denoising network parameters are optimized during training by a composite loss function. The first term is the conventional diffusion loss defined as:

\begin{equation}
\mathcal{L}_{\text{d}} = \mathbb{E}_{t,\mathbf{x},\epsilon} \left[\left\| \epsilon - \epsilon_\theta \left(\sqrt{\alpha_t} \mathbf{x} + \sqrt{1-\alpha_t}\epsilon, \mathbf{y}, t\right)\right\|_2\right],
\end{equation}

\noindent where $\epsilon$ is the true noise added during the forward process.

The model is trained to predict also the variances to improve the quality of generated images \cite{Quinn2021Improved}. To do so, the KL divergence between the predicted distribution $p_\theta(\mathbf{x}_{t-1} | \mathbf{x}_t, \mathbf{y})$ and the diffusion process posterior $q(\mathbf{x}_{t-1} | \mathbf{x}_t, \mathbf{x}_0)$ is added to the overall loss function as:

\begin{equation}
    \mathcal{L}_{\text{KL}} = \text{KL}(p_\theta(\mathbf{x}_{t-1} | \mathbf{x}_t, \mathbf{y}) \| q(\mathbf{x}_{t-1} | \mathbf{x}_t, \mathbf{x}_0)). 
\end{equation}

\noindent The resulting loss function is balanced by $\lambda$ as:

\begin{equation}
    \mathcal{L} = \mathcal{L}_{\text{d}} + \lambda \mathcal{L}_{\text{KL}}.
\end{equation}

\subsection{Quantization for Semantic Diffusion Model}

While post-training quantization (PTQ) can significantly reduce the memory footprint and computational complexity of a model by compressing neural networks by rounding weights to a discrete set of values, it introduces quantization noise to the weights of the well-trained network. This leads to quantization errors that may accumulate across layers \cite{dettmers2022llmint88bitmatrixmultiplication}, making deeper neural networks challenging to quantize.

In diffusion models, this problem is further exacerbated by the repeated nature of the denoising process. At each time step \( t \), the input to the denoising model is derived from the model output at the previous time step, resulting in a compounding of quantization errors as the process progresses. Thus, the biggest challenge of quantizing diffusion models is to find optimal quantization parameters that minimize the quantization error.

PTQ methods compress the models by rounding the weights $w$ to a discrete set of values following:

\begin{equation}
    \hat{w} = s \cdot \text{clip}(\text{round}(\frac{w}{s}), c_{\min}, c_{\max}),
\end{equation}

\noindent in which $s$ is the quantization scale, and $c_{\min}$ and $c_{\max}$ are the lower and upper bound of the clipping function set during the calibration process. To calibrate the quantized denoising network, we split the denoising U-Net into different blocks following the inner structure of the model that comprises several residual blocks. Then, we reconstruct outputs and finetune both the clipping parameters and the scaling factors of weight quantizers in each block with adaptive rounding \cite{Nagel2020UpOD, li2023qdiffusionquantizingdiffusionmodels}.

The proposed PTQ method introduces a timestep-aware calibration data sampling mechanism and a \textit{split} quantization technique, therefore quantizing the weights before the concatenation operation, to manage abnormal activation and weight distributions in shortcut layers, similar to \cite{li2023qdiffusionquantizingdiffusionmodels}. By adopting these strategies, we aim to mitigate the accumulation of quantization errors and preserve the model performance throughout the denoising process.


\subsection{Calibration Dataset Creation}
In applying post-training quantization (PTQ), it is essential to have a calibration dataset, which is significantly smaller than the training dataset for efficiency purposes, to capture activation values and optimize the quantization parameters. The distribution of the calibration samples should closely match the real data distribution to prevent overfitting the quantization parameters to the calibration set. Input noises for diffusion process are more constructive for calibrating quantized diffusion models \cite{shang2023posttrainingquantizationdiffusionmodels}. However, the variation in output activation distributions across time steps adds further complexity to the quantization process. To address this, similar to \cite{li2023qdiffusionquantizingdiffusionmodels}, we involve GESCO to randomly sample intermediate inputs uniformly over a fixed interval across all time steps to generate a calibration set representing the entire distribution. The reason behind this approach is that activation distributions change gradually over time, with the intermediate time steps covering a wide range of the distribution, while the distant endpoints differ more significantly. To further simulate training data and make the model fully aware of the conditions, our method generates the calibration dataset using as input semantic maps corrupted by additive white Gaussian noise (AWGN) at varying PSNR levels. Specifically, we consider PSNR values ranging from 1 to 100, uniformly distributing these noise intensities across the calibration dataset. This ensures that the calibration process accounts for the model robustness to noise, a key requirement in semantic communication scenarios.

\section{Experimental Evaluation}
\label{sec:exp}

\begin{figure}
    \centering
    \includegraphics[width=\linewidth]{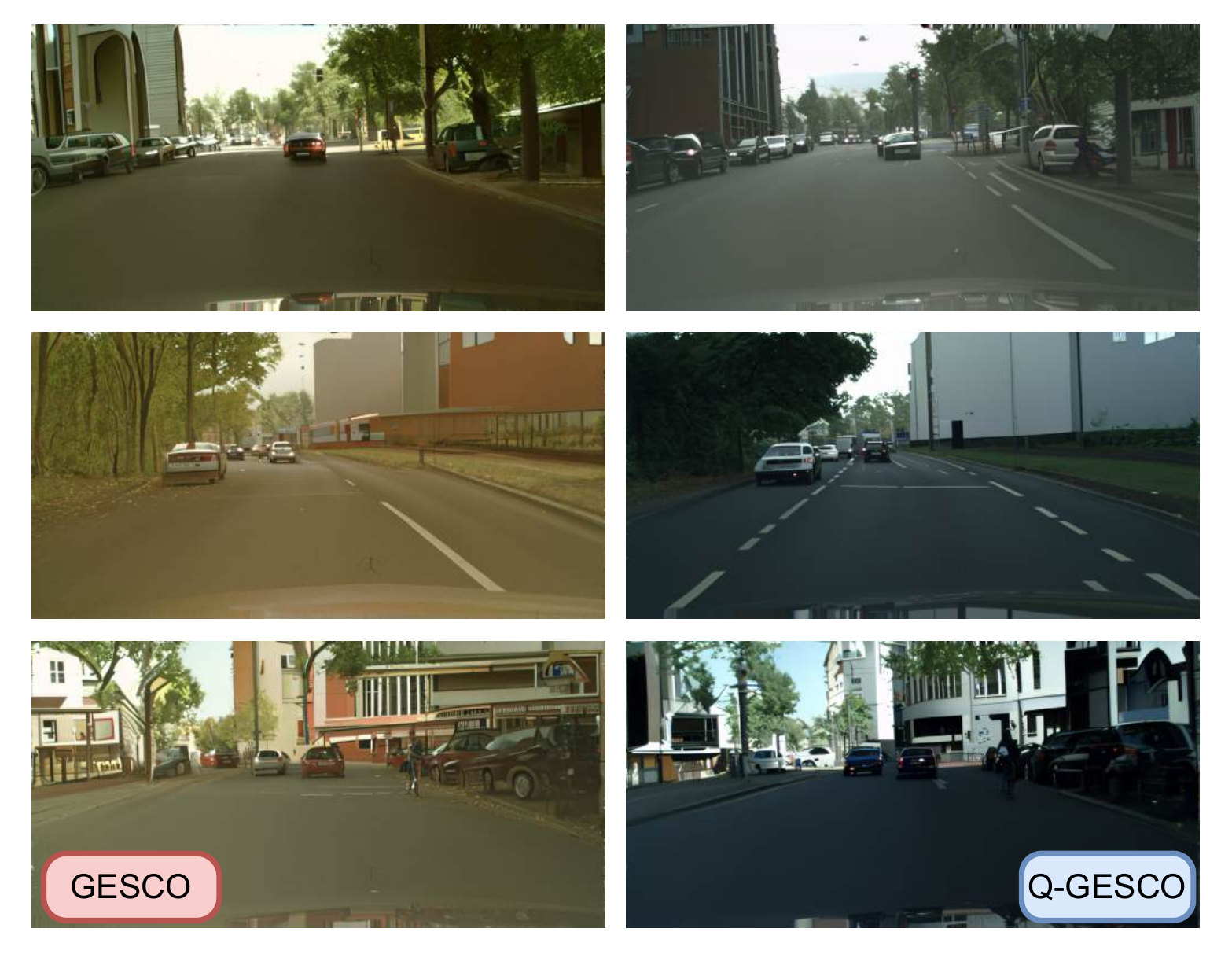}
    \caption{Sample results of Q-GESCO (right) compared to its full-precision counterpart GESCO (left) with PSNR = 10.}
    \label{fig:samples}
\end{figure}

\subsection{Experimental Setup}


\begin{table*}[]
\centering
\caption{Quantization results for the proposed Q-GESCO and its full-precision counterpart in terms of memory footprint (Bits, Size), computational load (FLOPs), and regenerated image quality (LPIPS, FID).}
\label{tab:results}
\begin{tabular}{@{}l|lll|cc|cc|cc@{}}
\toprule
& & & & \multicolumn{2}{c|}{PSNR 100} & \multicolumn{2}{c|}{PSR 20} & \multicolumn{2}{c}{PSR 10} \\
\toprule
Method  & Weight Bits           & Size (MB) & TFLOPs & LPIPS $\downarrow$ & FID ($\times10$) $\downarrow$ &                      LPIPS $\downarrow$ & FID ($\times10$) $\downarrow$ &  LPIPS $\downarrow$ & FID ($\times10$) $\downarrow$  \\ \midrule
GESCO   & 32                &  2699     & 13.82 &  0.65     & 16.43    & 0.66   & 17.10 & 0.67 & 20.29                   \\
                     Q-GESCO & 8 (\textcolor{red}{-75\%}) &  674 (\textcolor{red}{-75\%}) & 2.87 (\textcolor{red}{-79\%}) & \textbf{0.63}  & \textbf{15.82}  &    \textbf{0.65}   & \textbf{16.30} & \textbf{0.66} & \textbf{18.26}                \\ \bottomrule
\end{tabular}
\end{table*}

\textbf{Dataset.} We conduct the experiments on a well-known dataset in the literature, the Cityscapes dataset, consisting of video frames recorded in street scenes from different cities. This dataset provides images at a resolution of $256\times512$ of urban scene semantics with corresponding semantic maps for a total of 35 distinct classes. Cityscapes is employed both to construct the calibration dataset for post-training quantization and to generate samples for evaluating the quantized model.

\textbf{Communication Scenario.} The experimental setup repeats the communication scenario outlined in GESCO \cite{Grassucci2023GenerativeSC}. In this setup, the transmitter sends one-hot encoded semantic maps, which are compressed and normalized before being transmitted over a noisy communication channel. The receiver, upon receiving these noisy semantic maps, uses them to condition the denoising network to generate semantically equivalent images through the quantized diffusion model.

\textbf{Noise.} In line with \cite{Grassucci2023GenerativeSC}, we simulate the impact of the simple scenario of additive white Gaussian noise (AWGN) on the communication channel. In the experiments, we consider a range of Peak Signal-to-Noise Ratios (PSNR) between 0 and 100, allowing us to evaluate the robustness of the quantized model under varying levels of noise.

\textbf{Baseline.} The quantization process is applied to the model provided in GESCO \cite{Grassucci2023GenerativeSC}. We adopt the DDPM sampling with $T = 1000$ steps for the diffusion process. Additionally, the guidance scale $s$ was set to 2 to balance the trade-off between diversity and fidelity in the generated samples.

\textbf{Calibration Process.} The 64 samples for the calibration dataset are generated using the pre-trained GESCO model with DDIM, applying \( T = 100 \) steps for the inference process, starting from one-hot encoded semantic maps as conditioning. We select intermediate samples every 25 steps of the diffusion process to capture a representative range of activation distributions.
For channel noise, we uniformly distribute eight noise levels between 1 and 100 across the 64 generated samples. This approach is intended to increase the model robustness to noise, particularly in scenarios where the transmission is heavily affected by perturbations.
Our experiments execute the calibration process of the Q-GESCO model on an NVIDIA RTX8000 with 48 GB of VRAM. The process takes approximately 13 hours.

\subsection{Evaluation}

We conduct experiments to evaluate the memory footprint, the computational load, and the perceptual and visual quality of the regenerated images of Q-GESCO against the full-precision counterpart.
Table~\ref{tab:results} shows the objective metrics and the quantization results of the models. Q-GESCO saves 75\% of memory and 79\% of FLOPs with respect to the full-precision model, while obtaining comparable or superior generative performance.
To further analyze the generation ability of the proposed method, we conduct a visual inspection of the regenerated images. Figure~\ref{fig:samples} displays some samples generated at fixed PSNR = 10 for the full-precision GESCO and the proposed quantized framework Q-GESCO. Despite the reduced precision thanks to the quantization process, Q-GESCO still regenerates high-quality images even in the case of bad channel conditions, proving once again the effectiveness of the proposed method and the possibility to involve it in resource-constrained environments.


\subsection{Ablation Studies}

We evaluate the effectiveness of the noise-aware calibration process proposed in Sec.~\ref{sec:method} with an ablation study. Specifically, we first calibrate the quantized model with the timestep-aware only dataset, and then with the channel noise- and timestep-aware dataset to verify whether the simulation of the channel noise in the calibration process affects the performance. Table~\ref{tab:abl} shows the objective metrics evaluation of the quality of regenerated images. The proposed noise- and timestep-aware calibration process achieves the best results in terms of both LPIPS and FID, proving the success of the proposed calibration dataset receipt.

\begin{table}[]
\centering
\caption{Ablation studies for the calibration dataset.}
\label{tab:abl}
\begin{tabular}{@{}l|cc|cc@{}}
\toprule
               & \multicolumn{2}{c|}{PSNR 100} & \multicolumn{2}{c}{PSNR 10}  \\ 
\midrule 
Calibration Method   & LPIPS $\downarrow$ & FID ($\times10$) $\downarrow$ & LPIPS $\downarrow$ & FID ($\times10$) $\downarrow$ \\ \midrule
Timestep             & 0.69               & 16.95                       & 0.67               & 18.63                      \\
Noise + Timestep     & \textbf{0.63}      & \textbf{15.82}              & \textbf{0.66}      & \textbf{18.26}             \\ 
\bottomrule
\end{tabular}
\end{table}

\section{Conclusion}
\label{sec:con}

In this paper, we presented a novel generative semantic communication framework for resource-constrained environments. The proposed method exploits the abilities of the diffusion model to regenerate high-quality and semantic-preserving samples at the receiver while crucially reducing both the computational load and memory footprint of the inner neural network, enabling the usage of the proposed framework in resource-constrained semantic communication scenarios. We show that our method effectively reduces the demands of loading and memory with respect to the full-precision counterpart while preserving the generative performance in terms of both objective metrics and visual inspection evaluations.

\balance
\bibliographystyle{IEEEbib}
\bibliography{QGESCO}

\end{document}